\documentclass[prl,twocolumn,superscriptaddress,floatfix,nofootinbib]{revtex4-2}
\pdfoutput=1 
\usepackage[T1]{fontenc} 
\usepackage{amsmath,amssymb,amsfonts,amsthm}
\usepackage{graphicx}
\usepackage[usenames,dvipsnames]{color}
\usepackage{enumitem}
\usepackage{verbatim}
\usepackage[percent]{overpic}
\usepackage{rotating}
\usepackage{hyperref}
\usepackage{array}
\usepackage{mathtools}
\usepackage{lmodern}
\usepackage[normalem]{ulem}
\usepackage{braket}
\usepackage{tensor}
\usepackage{dsfont}
\usepackage{bm}
\usepackage{tikz}
\usepackage{bbm}

\newcommand{\EndMatter}{
\vskip 0.5cm
\centerline{\textbf{END MATTER}}
\vskip 0.5cm
}

\usepackage{mathrsfs}

\usetikzlibrary{decorations.pathmorphing}
\usepackage{CJKutf8}

\usepackage[whole]{bxcjkjatype}

\usetikzlibrary{positioning}
\usetikzlibrary{calc,through,backgrounds}

\theoremstyle{definition}
\newtheorem{definition}{Definition}[section]
\newtheorem{theorem}{Theorem}[section]

\newcommand{\kako}[1]{\left( #1 \right)}
\newcommand{\kagikako}[1]{\left[ #1 \right]}
\newcommand{\ts}[1]{ _{\text{#1}} }

\allowdisplaybreaks

\DeclareMathOperator{\Tr}{Tr}
\newcommand{\R}{\mathbb{R}}
\newcommand{\C}{\mathbb{C}}
\newcommand{\dd}{\text{d}}

\newcommand{\id}{\mathbbm{1}}
\newcommand{\sx}{\mathsf{x}}

\newcommand{\ii}{\mathrm{i}}

\newcommand{\mfd}{\mathcal{M}}

\newcommand{\opeU}{\hat{\mathcal{U}}}
\newcommand{\retG}{G_{\mathrm{R}}}
\newcommand{\Var}{\mathrm{Var}}
\newcommand{\scri}{\mathscr{I}}

\newcommand{\renorm}[1]{:\!\!#1\!\!:}

\newcommand{\solR}{\mathrm{Sol}_{\R}(\mfd)}

\newcommand{\dvol}[1]{\mathrm{d}\mathsf{V}_{#1}}
\newcommand{\ave}[1]{\mathbb{E}\kagikako{#1}}
\newcommand{\var}[1]{\mathbb{V}\left[#1\right]}
\newcommand{\form}[1]{\bm{\mathrm{#1}}}

\begin{document}

\title{Trade-off Relation for Black Hole Entropy Fluctuations}

\author{Kensuke Gallock-Yoshimura}
\email{gallockyoshimura@biom.t.u-tokyo.ac.jp}

\affiliation{
Department of Electrical Engineering and Information Systems, Graduate School of Engineering, \\
The University of Tokyo, 
7–3–1 Hongo, Bunkyo-ku, Tokyo 113–8656, Japan
}

\author{Yoshihiko Hasegawa}
\email{hasegawa@biom.t.u-tokyo.ac.jp} 

\affiliation{
Department of Electrical Engineering and Information Systems, Graduate School of Engineering, \\
The University of Tokyo, 
7–3–1 Hongo, Bunkyo-ku, Tokyo 113–8656, Japan
}

\begin{abstract}
Black holes respond to infalling quantum matter fields by changing their entropy. 
Since such matter is quantum in nature, the entropy response should be sensitive to its quantum fluctuations. 
We show, within stochastic semiclassical gravity, that a horizon cannot record relevant quantum information with arbitrarily small entropy fluctuations. 
For the infalling photons encoding which-path information in the Danielson-Satishchandran-Wald decoherence experiment, we derive a trade-off relation between the stochastic variance of the black hole entropy change and the photon number. 
\end{abstract}

\maketitle
\flushbottom

\emph{Introduction}---%
Much of our understanding of black holes, such as their entropy \cite{Bekenstein.BH2ndLaw.1972, Bekenstein.BHentropy.1973, Bekenstein.GSL.1974}, the first law of black hole thermodynamics \cite{wald1994QFT, Sijie.physical.process.2001, Hollands.entropyBH.2024}, and the fate of quantum information \cite{tHooft.dimensional.1993, Susskind.hologram.1995, Susskind.BHcomplementarity.1993, Hayden.Preskill.2007, AMPS.2013}, comes from examining their response to matter crossing the horizon. 
Yet, this response is most often formulated semiclassically. 
In semiclassical gravity, geometry reacts to the expectation value of the quantized matter stress tensor, and the corresponding black hole entropy response is characterized by its mean change. 
However, if the matter entering the horizon is quantum mechanical, the associated stress tensor carries quantum fluctuations. 
Such matter should induce fluctuations in geometry, and hence in the black hole entropy beyond the semiclassical framework. 
It is then natural to ask not only how black hole entropy responds on average, but also how deterministic it can be.

In this work, we show that black hole horizons cannot record infalling quantum information with arbitrarily small entropy fluctuations. 
Within \emph{stochastic semiclassical gravity}, where geometry obeys the Einstein-Langevin equations and responds both to the expectation value and to quantum fluctuations of the matter stress tensor (see \cite{Hu.stochastic.primer.2003, Hu.StochasticGravity.2008} for a review), we derive a trade-off relation \eqref{eq:BH TUR} among the stochastic mean of the change $\mathbb E[\Delta \delta S]$ in the perturbed black hole entropy $\delta S$, its variance $\mathbb V[\Delta \delta S]$, and the number of photons $\braket{\hat N}$ carrying the relevant quantum information into the horizon: $\mathbb V[\Delta \delta S]/(\mathbb E[\Delta \delta S])^2 \geq 1/(4 \braket{\hat N})$. 
Thus, for a given number of information-carrying photons $\braket{\hat N}$, the change in black hole entropy has an unavoidable fluctuation.

\begin{figure}[t]
\centering
\includegraphics[width=\linewidth]{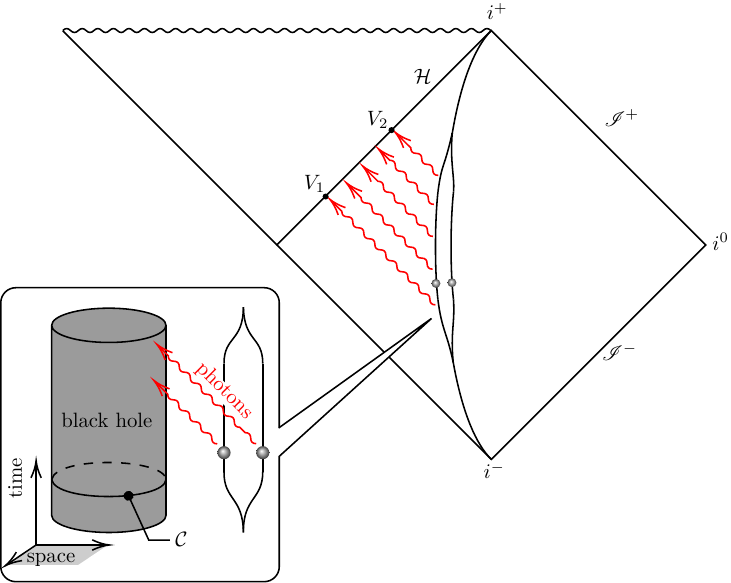}
\caption{The Penrose diagram of a black hole with stationary trajectories of the superposed charge and a schematic picture of the DSW experiment. 
No radiation will be propagated to $\scri^+$ so long as the experiment is performed adiabatically. 
Here, $V_1$ and $V_2$ are two ``times'' on the horizon $\cal H$, and $\cal C$ is a cross section of the black hole. 
}
\label{fig:DSWdiagram}
\end{figure}

A key ingredient for deriving the trade-off relation is the black hole-induced decoherence mechanism found by Danielson, Satishchandran, and Wald (DSW) \cite{DSW.BH.2022, DSW.Killing.2023} (see also \cite{Gralla.rotatingBH.2024, DSW.Local.2025, Li.RNBH.2025, DKSW.minimize.2025}). 
In the DSW mechanism, the horizon memory effect inevitably destroys the quantum coherence of a spatially superposed charge outside the black hole by capturing photons entangled with the charge's center-of-mass degree of freedom (Fig.~\ref{fig:DSWdiagram}). 
In other words, black holes acquire ``which-path information'' about the superposed trajectory encoded in the entangling photons. 
The photon number $\braket{\hat N}$ in our trade-off relation is therefore not an externally prescribed resource, but rather it quantifies a black hole's ability to record which-path information. 
Therefore, the resulting trade-off relates the black hole entropy fluctuations to the intrinsic ability of a horizon to record quantum information.

\emph{Dephasing structure of the DSW mechanism}---%
To obtain the trade-off relation \eqref{eq:BH TUR}, we consider a quantum flux generated by the DSW mechanism and derive a lemma relating the variance of the Killing energy of the entangling photons to the photon number, \eqref{eq:main result for charge}. 
We then apply this to the infalling quantum flux perturbing an asymptotically flat and stationary black hole geometry with vanishing charge and angular momentum in stochastic semiclassical gravity.

We first isolate the essential features of the DSW mechanism that are needed for our trade-off relation. 
The full DSW setup involves a spatially superposed charged particle coupled to a quantized electromagnetic field in a black hole spacetime. 
However, for the purpose of deriving the trade-off relation, only four structural ingredients are essential: (i) the two branches of the superposition constitute an effective two-level system \cite{WilsonGerow.warmHorizon.2024}; 
(ii) the electromagnetic four-potential $\hat A_\mu$ is the vector field analogue of the Klein-Gordon scalar field \cite{wald1994QFT}; 
(iii) the DSW decoherence is of pure-dephasing type and the emitted photons are multimode coherent state entangled with the charge's superposition state \cite{DSW.BH.2022, DSW.Killing.2023}; and 
(iv) the electromagnetic stress-energy tensor $\hat T_{\mu \nu}= \hat F_{\mu}^{\phantom{\mu}\lambda} \hat F_{\nu \lambda} - \frac14 g_{\mu \nu} \hat F_{\alpha \beta} \hat F^{\alpha \beta}$, where $\hat F_{\mu \nu}=2 \nabla_{[\mu}\hat A_{\nu]}$ is the field strength, has the same structural form as in the scalar case. 
These ingredients are captured by a simple model described by a dephasing quantum system locally coupled to a quantum scalar field. 
For this reason, we employ such a simplified model and derive our trade-off relation.


Let $(\mfd, g_{\mu \nu})$ be a $(3+1)$-dimensional globally hyperbolic, stationary spacetime with a timelike Killing vector field $\xi^\mu$. 
We consider the spacetime-smeared real Klein-Gordon field $\hat \phi(f)\coloneqq \int_\mfd \dvol{g}\,f(\sx) \hat \phi(\sx)$, where $\dvol{g} \equiv \sqrt{-\det g_{\mu \nu}} \dd^{4} \sx$ is the invariant volume element and $f \in C_0^\infty (\mfd)$ is a compactly-supported smooth test function on $\mfd$. 
Such a field observable is assumed to satisfy the Klein-Gordon equation, $(\nabla_\mu \nabla^\mu - W) \hat \phi(f)=0$, hermiticity $\hat \phi(f)^\dag = \hat \phi(f)$, and the commutation relation $[\hat \phi(f), \hat \phi(f')]=\ii E(f,f')\id$. 
Here, $W$ is a smooth function on $\mfd$ and $E(f,f')\coloneqq \int_{\mfd\times \mfd} \dvol{g} \dvol{g'}\,f(\sx) f(\sx') E(\sx, \sx')$ is the smeared causal propagator.

Throughout this work, we take the field state $\rho\ts{F}$ to be quasifree and Hadamard. 
Quasifree states, such as vacuum, thermal, and squeezed states, have vanishing one-point correlator, $\braket{\hat \phi(f)}\equiv \Tr[ \rho\ts{F} \hat \phi(f) ]=0$, and any $n(>2)$-point correlator $\braket{\hat \phi(f_1)\ldots \hat \phi(f_n)}$ can be expressed in terms of products of two-point correlators $\braket{\hat \phi(f)\hat \phi(f')}$. 
The Hadamard condition fixes the short-distance behavior of the two-point correlator, which is essential here because it permits renormalization of the stress-energy tensor, $\hat T_{\mu \nu}(\sx)=\nabla_{(\mu} \hat \phi \nabla_{\nu)}\hat \phi - \frac12 g_{\mu \nu} (\nabla^\lambda \hat \phi \nabla_\lambda \hat \phi + W\hat \phi^2)$. 
We will write the normal-ordered stress-energy tensor as $\renorm{\hat T_{\mu \nu}} = \hat T_{\mu \nu} - \braket{\hat T_{\mu \nu}}_0 \id$, where $\braket{\hat T_{\mu \nu}}_0 \equiv \Tr[\rho\ts{F,0} \hat T_{\mu \nu}]$ is the expectation value with respect to a quasifree Hadamard reference state $\rho\ts{F,0}$. 
Since any Hadamard state $\rho\ts{F,0}$ on a Cauchy slice $\Sigma_0$ evolves to Hadamard $\rho\ts{F}$ on another Cauchy slice $\Sigma_t$, the expectation value, $\braket{\renorm{\hat T_{\mu \nu}}} \equiv \Tr[\rho\ts{F} \renorm{\hat T_{\mu \nu}}] =\braket{\hat T_{\mu \nu}} - \braket{\hat T_{\mu \nu}}_0$, is free of UV divergences on any Cauchy slice. 
For a spacetime with a timelike Killing vector field $\xi^\mu$, the normal-ordered conserved energy $\renorm{\hat Q[\xi]} \coloneqq \int_\Sigma \dd \Sigma^\mu \renorm{\hat T_{\mu \nu}} \xi^\nu$ also gives a UV-finite expectation value, $\braket{\renorm{\hat Q[\xi]}}$.

We now consider a localized $d$-level quantum system (described by the Hilbert space $\C^d$) coupled to a quantum scalar field on a general $(3+1)$-dimensional globally hyperbolic spacetime $\mfd$. 
To connect with the DSW decoherence scenario, we employ the pure-dephasing model \cite{Tjoa.Nonperturbative.2023}, which captures the essential features of the DSW decoherence. 
The pure-dephasing model can be generically described by the total Hamiltonian, 
\begin{align}
    \hat H\ts{tot}
    &=
        \hat H\ts{sys}
        +
        \hat H\ts{F}
        + 
        \hat S \otimes \hat \phi(f)\,, \label{eq:total Hamiltonian}
\end{align}
where $\hat H\ts{sys}$ and $\hat H\ts{F}$ are the free Hamiltonians of the localized quantum system and the field, respectively, and $\hat S \in \mathcal B(\C^d)$ is a bounded hermitian operator for the system that commutes with the free Hamiltonian, $[\hat H\ts{sys}, \hat S]=0$. 
We will assume that $\hat S$ has a finite discrete spectrum $\{ s \}$ with eigenvectors $\{ \ket{s} \}$. 
The test function $f\in C_0^\infty (\mfd)$ corresponds to the interaction region. 
Since the free Hamiltonians commute with the interaction Hamiltonian, the dynamics is purely dephasing, i.e., populations are conserved while coherences decay exponentially. 
Such a pure dephasing model is not only suitable for analyzing the DSW decoherence, but also it enables us to employ a nonperturbative analysis. 
To see this, we consider the dephasing model in the interaction picture. 
The interaction Hamiltonian density is $\hat h\ts{I}(\sx)=\hat S \otimes f(\sx) \hat \phi(\sx)$, and the time-evolution unitary operator $\hat U$ is given by the Magnus expansion \cite{Blanes.Magnus.2009} in a coordinate-independent form as $\hat U=\exp (\sum_{k=1}^\infty \hat{\mathcal U}_k)$ with 
\begin{subequations}
\begin{align}
    \hat{\mathcal U}_1
    &\coloneqq
        -\ii \int_\mfd \dvol{g}\,\hat h\ts{I}(\sx)\,, \\
    \hat{\mathcal U}_2
    &\coloneqq
        -\dfrac{1}{2}
        \int_\mfd \dvol{g}
        \int_\mfd \dvol{g'}\,
        \Theta(t-t') 
        [\hat h\ts{I}(\sx), \hat h\ts{I}(\sx')]\,, 
\end{align}
\end{subequations}
where $\Theta(t)$ is the Heaviside step function for an arbitrary time parameter $t$. 
All higher order terms $\hat{\mathcal U}_{k\geq 3}$ vanish since the field commutator is a c-number, $[\hat \phi(\sx), \hat \phi(\sx')] \in \C$ \cite{Landulfo2016magnus1, Cozzella.UDW.2020, Gallock.Acceleration.2025}. 
Defining the retarded Green function $\retG(\sx, \sx')\equiv -\ii \Theta(t-t') [ \hat \phi (\sx), \hat \phi(\sx') ]$ and the spacetime-smeared retarded Green function $\retG(f, f)$ as 
\begin{align}
    &\retG(f, f)
    \equiv 
        \int_\mfd \dvol{g}
        \int_\mfd \dvol{g'}
        f(\sx)
        f(\sx')
        \retG(\sx, \sx')\,,
\end{align}
the time-evolution unitary operator reduces to a simpler form
\begin{align}
    \hat U
    &=
        e^{\opeU_1}
        e^{\opeU_2}\,, \label{eq:unitary operator}
\end{align}
where 
\begin{subequations}
\begin{align}
    \opeU_1
    &=
        -\ii \hat S \otimes \hat \phi(f)
    =
        -\hat S \otimes [ \hat a^\dag(KEf) - \hat a(\overline{KEf}) ]\,, \\
    \opeU_2
    &=
        -\dfrac{\ii}{2}
        \retG(f, f) \hat S^2 \otimes \id\,.
\end{align}
\end{subequations}
Here, $KEf$ is the positive frequency mode and  $\overline{KEf}$ is its complex conjugate \cite{wald1994QFT}, $\hat a(\overline{KEf})$ and $\hat a^\dag(KEf)$ are respectively the annihilation and creation operators such that $\hat a(\overline{KEf}) \ket{\Omega_0}=0$ for all $f\in C_0^\infty (\mfd)$ for a given vacuum state $\ket{\Omega_0}$ of the field. 
The time-evolved state of an initially separable state $\ket{\psi\ts{sys,0}} \otimes \ket{\Omega_0}$ with $\ket{\psi\ts{sys,0}}=\sum_s \psi_s \ket{s}$ is \cite{Cozzella.UDW.2020, Gallock.Acceleration.2025}
\begin{align}
    &\ket{\Psi\ts{fin}}
    =
        \hat U \ket{\psi\ts{sys,0}} \ket{\Omega_0} \notag \\
    &=
        \sum_s \psi_s e^{-\ii \retG(f,f) s^2/2} \ket{s}
        \otimes \hat D(-s KEf) \ket{\Omega_0}\,, \label{eq:final state}
\end{align}
where $\hat D(-s KEf) \equiv \exp [ \hat a^\dag (-sKEf) - \hat a(-\overline{sKEf}) ]$ is the displacement operator. 
Thus, the pure-dephasing quantum system entangles with the field in a (multimode) coherent state.

\emph{Bounds for boson number and energy}---%
We compute the variances of the entangling photon number and the conserved energy of the quantum field in the pure-dephasing model and derive lower bounds on them. 
The resulting bounds also apply to the quantum flux generated through the DSW mechanism.  
Consider an initially separable state, $\rho\ts{sys,0}\otimes \ket{\Omega_0}\bra{\Omega_0}$, where $\rho\ts{sys,0}=\sum_{s,s'} \rho_{ss'}\ket{s}\bra{s'}$ and $\ket{\Omega_0}$ is the quasifree Hadamard reference state. 
Using the final state \eqref{eq:final state}, the expectation value of the number of bosons after the interaction reads $\braket{\hat N} \equiv \Tr\ts{F}[\rho\ts{F} \hat N]= \mathcal N \braket{\hat S^2}$, where 
\begin{align}
    \mathcal N 
    =
        \braket{\hat \phi(f) \hat \phi(f)}_0
    \equiv 
        \braket{\Omega_0| \hat \phi(f) \hat \phi(f) |\Omega_0}
    \label{eq:degree of decoherence}
\end{align}
is the smeared vacuum two-point correlator and $\braket{\hat S^2}= \sum_s \rho_{ss} s^2$. 
The quantity $\cal N$ determines the strength of decoherence, since it exponentially suppresses the off-diagonal elements of the system \cite{Gallock.Acceleration.2025}, $\rho\ts{sys}=\sum_{s,s'} \rho_{s,s'} e^{-\ii \retG(f,f) (s^2 - s^{\prime 2})/2} e^{ -(s-s')^2 \mathcal N/2 } \ket{s}\bra{s'}$. 
The variance of the number operator, $\Var[\hat N]\coloneqq \braket{\hat N^2} - \braket{\hat N}^2$, can be straightforwardly computed as $\Var[\hat N]=\braket{\hat N} + \braket{\hat N}^2 \Var[\hat S^2] /\braket{\hat S^2}^2$. 
The second term on the right-hand side is non-negative and depends on the system's initial state. 
Thus, the boson number fluctuations obey the universal bound 
\begin{align}
    \Var[\hat N] \geq \braket{\hat N}\,. \label{eq:number variance ineq}
\end{align}
Since $\braket{\hat N}$ is tied to decoherence $\cal N$, the inequality \eqref{eq:number variance ineq} tells us that stronger decoherence implies larger number fluctuations. 
The equality holds if $\Var[\hat S^2]=0$, such as $\hat S=\hat \sigma_z$ for $d=2$.

In the same manner, we compute the variance of the renormalized conserved energy, $\Var[\renorm{\hat Q[\xi]}]$. 
To this end, we first introduce the following bilinear form, 
\begin{align}
    T_{\mu \nu}[\varphi_1, \varphi_2]
    &\coloneqq
        \nabla_{(\mu} \varphi_1 
        \nabla_{\nu)} \varphi_2
        -\dfrac{1}{2}
        g_{\mu \nu} 
        (
            \nabla^\lambda \varphi_1 \nabla_\lambda \varphi_2
            + 
            W \varphi_1 \varphi_2
        )
        \,,
\end{align}
where $\varphi_1, \varphi_2 \in \solR$ are the real solutions to the Klein-Gordon equation. 
The stress-energy operator thus can be expressed as $\hat T_{\mu \nu}(\sx)=T_{\mu \nu}[\hat \phi, \hat \phi]$. 
The Baker-Campbell-Hausdorff formula tells us that 
\begin{align}
    \hat U^\dag \hat T_{\mu \nu} \hat U
    &=
        \hat T_{\mu \nu}
        + 
        \hat S \otimes T_{\mu \nu}[\hat \phi, Ef]
        + 
        \mathcal T_{\mu \nu} \hat S^2\,, \label{eq:stress tensor evolution}
\end{align}
where $Ef\equiv Ef(\sx)=\int_\mfd \dvol{g'}\,E(\sx, \sx') f(\sx') \in \solR$ is the advanced-minus-retarded real solution to the Klein-Gordon equation, and $\mathcal T_{\mu \nu} \equiv T_{\mu \nu}[Ef, Ef]$. 
By noting that $\braket{\hat \phi}_0=0$ as the field is initially quasifree, it is straightforward to obtain the expectation value of the renormalized stress-energy tensor as $\braket{\renorm{\hat T_{\mu \nu}}}=\braket{\hat S^2} \mathcal T_{\mu \nu}$, and therefore, the expectation value $\braket{\renorm{\hat Q[\xi]}}$ reads 
\begin{align}
    \braket{\renorm{\hat Q[\xi]}}
    &=
        \braket{\hat S^2} 
        \int_{\Sigma} \dd \Sigma^\mu\,
        \mathcal T_{\mu \nu} \xi^\nu\,. \label{eq:charge expactation}
\end{align}
A similar calculation gives the variance of $\renorm{\hat Q[\xi]}$ (see End Matter),  
\begin{align}
    &\Var[\renorm{\hat Q[\xi]}]
    =
        \Var_0[\renorm{\hat Q[\xi]}]
        +
        \dfrac{\Var[\hat S^2]}{\braket{\hat S^2}^2}
        \braket{\renorm{\hat Q[\xi]}}^2 \notag \\
    &\quad
    +
    \braket{\hat S^2}
    \int_{\Sigma} \dd \Sigma^\mu 
    \int_{\Sigma} \dd \Sigma^\rho\,
    \xi^\nu \xi^\sigma 
    \braket{ T_{\mu \nu}[\hat \phi, Ef] T_{\rho \sigma}[\hat \phi, Ef]}_0, \label{eq:varQ intermediate}
\end{align}
where $\Var[\bullet]_0 \equiv \braket{\bullet^2 }_0 - \braket{\bullet}^2_0$ is the variance with respect to the reference state $\rho\ts{F,0}=\ket{\Omega_0}\bra{\Omega_0}$. 
The key observation is that the integrand in the last term in \eqref{eq:varQ intermediate} can be thought of as a two-point correlation function on a Cauchy surface $\Sigma$. 
To see this, observe that 
\begin{align}
    \hat \Phi (h_f)
    \equiv 
        \int_\Sigma \dd \Sigma^\mu \,
        T_{\mu \nu } [\hat \phi, Ef] \xi^\nu
\end{align}
is a linear operator of $\hat \phi$, where $h_f$ is a corresponding function that smears $\hat \phi$ in the integrand. 
Therefore, the last term in \eqref{eq:varQ intermediate} is equivalent to $\braket{\hat S^2} \braket{\hat \Phi(h_f) \hat \Phi(h_f)}_0$. 
We then use the Cauchy-Schwarz inequality 
\begin{align}
    \braket{\hat \Phi(h_f) \hat \Phi(h_f)}_0 
    \braket{\hat \phi(f) \hat \phi(f)}_0 
    &\geq 
        | \braket{\hat \Phi(h_f) \hat \phi(f)}_0 |^2 \notag \\
    &\geq 
        |\mathrm{Im} \braket{\hat \Phi(h_f) \hat \phi(f)}_0|^2\,, \label{eq:Cauchy-Schwarz}
\end{align}
and we find $\mathrm{Im} \braket{\hat \Phi(h_f) \hat \phi(f)}_0= \int_\Sigma \dd \Sigma^\mu\, \mathcal T_{\mu \nu} \xi^\nu /2$. 
Thus, Eq.~\eqref{eq:varQ intermediate} becomes 
\begin{align}
    \dfrac{
        \Var[\renorm{\hat Q[\xi]}]
        -
        \Var_0[\renorm{\hat Q[\xi]}]
    }{\braket{\renorm{\hat Q[\xi]}}^2}
    \geq 
        \dfrac{ 1 }{ 4\braket{\hat N} }\,, \label{eq:main result for charge}
\end{align}
where we used Eq.~\eqref{eq:charge expactation}, $\braket{\hat N}=\braket{\hat S^2} \mathcal N$ [with Eq.~\eqref{eq:degree of decoherence}], and $\braket{\renorm{\hat Q[\xi]}}^2 \Var[\hat S^2]/\braket{\hat S^2}^2 \geq 0$. 
Equation \eqref{eq:main result for charge} is the central lemma from which our main result \eqref{eq:BH TUR} is derived. 
It represents a trade-off relation between the number of entangling photons, the field energy, and its fluctuation. 
Inequalities of this type (a variance normalized by the squared mean is bounded from below) have attracted significant interest in stochastic thermodynamics \cite{Barato.TUR.2015, Gingrich.Dissipation.2016, Hasegawa.QTUR.measurement.2020, Hasegawa.QTURgeneral.2021}.

\emph{Trade-off relation for black hole entropy}---%
We now apply our previous result to a black hole spacetime in the stochastic semiclassical gravity framework. 
In particular, we consider the infalling entangling photons generated by the DSW mechanism \cite{DSW.BH.2022, DSW.Killing.2023}, whose Killing energy satisfies the inequality \eqref{eq:main result for charge}. 
We then analyze their backreaction on the gravitational field as an external matter perturbation \cite{Hollands.entropyBH.2024}.

We begin with an asymptotically flat and stationary unperturbed black hole geometry with vanishing charge and angular momentum in the stochastic semiclassical gravity framework. 
Before the DSW experiment, the quantum field is in the quasifree, Hadamard reference state $\rho\ts{F,0}$. 
Due to the stationarity of the unperturbed geometry, we assume $\rho\ts{F,0}$ does not contribute to the geometry, i.e., $\braket{\renorm{\hat Q[\xi]}}_0=0$ and $\Var_0[\renorm{\hat Q}]=0$. 
The unperturbed black hole then satisfies the vacuum field equations, $(E\ts{G})_{\mu \nu}=0$.

An exterior observer, Alice, then performs the DSW decoherence experiment adiabatically. 
Within stochastic semiclassical gravity, the first-order gravitational field equations for the metric perturbation $\delta g$ sourced by the infalling flux are given by the Einstein-Langevin equations \cite{Hu.stochastic.primer.2003, Hu.StochasticGravity.2008}: 
\begin{align}
    2 \delta (E\ts{G})_{\mu \nu}
    =
        \braket{ \renorm{\hat T_{\mu \nu}} }
        +
        \tau_{\mu \nu}\,,
\end{align}
where $\tau_{\mu \nu}$ is a Gaussian stochastic tensor field associated with the infalling matter satisfying 
\begin{align}
    &\ave{\tau_{\mu \nu}(\sx)}=0\,, \\
    &\var{\tau_{\mu \nu}(\sx) \tau_{\rho \sigma}(\sx')}
    \equiv 
        N_{\mu \nu \rho \sigma}(\sx, \sx')
    \coloneqq
        \dfrac{1}{2} 
        \braket{
             \{ 
                \hat t_{\mu \nu}(\sx), \hat t_{\rho \sigma}(\sx')
             \}
        }.
\end{align}
Here, $\ave{\bullet}$ and $\var{\bullet}$ denote the stochastic mean and variance with respect to $\tau_{\mu \nu}$, defined by the path integral $\ave{X}\coloneqq \int \mathcal D\tau\,P[\tau] X[\tau]$, where $P[\tau] \propto \exp[-\int \tau_{\mu \nu}(N^{-1})^{\mu \nu \rho \sigma} \tau_{\rho \sigma}/2 ]$ is the probability functional, and $N_{\mu \nu \rho \sigma}(\sx, \sx')$ is the noise kernel bi-tensor. 
The operator $\hat t_{\mu \nu}$ is given by $\hat t_{\mu \nu}(\sx) \coloneqq \hat T_{\mu \nu}(\sx) - \braket{\hat T_{\mu \nu}(\sx)} \id$, where the expectation value is understood as $\braket{\hat T_{\mu \nu}} \equiv \Tr[\rho\ts{F} \hat T_{\mu \nu}]$. 
The semiclassical equations are recovered by taking the stochastic mean: $2\ave{\delta (E\ts{G})_{\mu \nu}}= \braket{\renorm{\hat T_{\mu \nu}}}$.

We now derive the black hole entropy trade-off relation by first realizing that the integral of the noise kernel on a Cauchy surface $\Sigma$ yields
\begin{align}
    \int_\Sigma \dd \Sigma^\mu \xi^\nu
    \int_{\Sigma'} \dd \Sigma^{\prime \rho} \xi^{\prime \sigma}
    N_{\mu \nu \rho \sigma}(\sx, \sx') 
    &=
        \Var[\renorm{\hat Q[\xi]}]
        \,.\label{eq:noise kernel and variance}
\end{align}
Crucially, no memory occurs at the future null infinity $\scri^+$ (Fig.~\ref{fig:DSWdiagram}), thereby all the photons propagate towards the horizon so long as Alice's interference experiment is conducted adiabatically \cite{DSW.BH.2022}. 
Therefore, taking the limit $\Sigma \to \mathcal H \cup i^+ \cup \scri^+$, Eq.~\eqref{eq:noise kernel and variance} becomes an equality on the event horizon $\cal H$. 
We shall assume that the energy flux due to such an experiment reaches the horizon $\cal H$ during ``times'' $V_1$ and $V_2(> V_1)$, where $V$ is an affine parameter on the horizon. 
The cross section of the black hole at each time is denoted by $\mathcal C_1 \equiv \mathcal C(V_1)$ and $\mathcal C_2 \equiv \mathcal C(V_2)$, and the portion of the horizon between these times by $\mathcal H_{12}$.

We employ the gravitational phase space method developed in \cite{Iyer.Noether.1994, Wald.Zoupas.2000, Hollands.stability.2012, Hollands.entropyBH.2024} and introduce the black hole's dynamical entropy. 
As reviewed in the Supplemental Material \cite{Supplemental}, the black hole entropy $S[\mathcal C]$ of a cross section $\cal C$ is defined as $S[\mathcal C]\coloneqq \int_{\mathcal C} \form{S}$, where $\form S$ is the entropy 2-form on the horizon $\mathcal H$ introduced in \cite{Hollands.entropyBH.2024}. 
In stochastic semiclassical gravity, the first-order variation of the entropy 2-form $\delta \form S$ on the horizon can be shown to satisfy \cite{Hollands.entropyBH.2024, Supplemental}
\begin{align}
    \dfrac{\kappa}{2\pi} \dd \delta \bm{\mathrm{S}}
    = 
        \bm{\mathrm{e}}
        +
        \bm \eta\,,
\end{align}
where $\kappa$ is the surface gravity, `$\dd$' is the exterior derivative and $\form e$ is the perturbed matter energy flux 3-form \cite{Hollands.entropyBH.2024} for which $e_{a_1 a_2 a_3}=-\xi^\mu \braket{\renorm{\hat T_\mu{}^\nu}} \epsilon_{\nu a_1 a_2 a_3}$ with $\form \epsilon$ the volume 4-form, and its integral over the horizon reads $\braket{\renorm{\hat Q[\xi]}}=\int_{\mathcal H_{12}} \form e$. 
The quantity $\form \eta$ is what we shall call the stochastic perturbed matter energy flux 3-form, for which $\eta_{a_1 a_2 a_3}=-\xi^\mu \tau_{\mu}{}^\nu \epsilon_{\nu a_1 a_2 a_3}$. 
Note that the physical process version of the first law of black hole thermodynamics \cite{wald1994QFT, Sijie.physical.process.2001, Hollands.entropyBH.2024} can be obtained by taking the stochastic mean $\mathbb E$ followed by the integral on the event horizon: 
\begin{align}
    \dfrac{\kappa}{2\pi} \Delta \ave{ \delta S }
    =
        \braket{\renorm{\hat Q[\xi]}}\,, \label{eq:1st law}
\end{align}
where $\Delta \ave{\delta S}\equiv \ave{\delta S[\mathcal C_2]} - \ave{\delta S[\mathcal C_1]}$ is the difference between the stochastic mean of perturbed black hole entropies of two cross sections $\mathcal C_1$ and $\mathcal C_2$ \cite{Hollands.entropyBH.2024}. 
By using the first law \eqref{eq:1st law} and the fact $N_{\mu \nu \rho \sigma}(\sx, \sx')=\ave{ \bm \eta(\sx) \bm \eta(\sx') }$, it is straightforward to show that \eqref{eq:noise kernel and variance} evaluated on the horizon is 
\begin{align}
    \Var[\renorm{\hat Q[\xi]}]
    &=
        \ave{
            \int_{\mathcal H_{12}} \bm \eta(\sx)
            \int_{\mathcal H_{12}} \bm \eta(\sx')
        } \notag \\
    &=
        \var{
            \dfrac{\kappa}{2\pi} \Delta \delta S
        }
        +
        \kako{
            \dfrac{\kappa}{2\pi} 
            \ave{
                \Delta \delta S
            }
            -
            \braket{\renorm{\hat Q[\xi]}}
        }^2 \notag \\
    &=
        \var{
            \dfrac{\kappa}{2\pi} \Delta \delta S
        }\,.\label{eq:Var S intermediate}
\end{align}
Applying our lemma \eqref{eq:main result for charge} with $\Var_0[\renorm{\hat Q[\xi]}]=0$ and the first law \eqref{eq:1st law}, we obtain the trade-off relation for the change in black hole entropy: 
\begin{align}
    &
    \dfrac{
        \mathbb V
        \kagikako{
            \Delta \delta S
        }
    }
    {
        \ave{ \Delta \delta S }^2
    }
    \geq 
        \dfrac{1}{4 \braket{\hat N}}\,. \label{eq:BH TUR}
\end{align}

Equation~\eqref{eq:BH TUR} is our main result. 
It states that the relative variance---variance divided by mean squared---of the change in the perturbed black hole entropy due to the infalling entangling photons is bounded from below by the inverse of the photon number, which originates from the memory effect on the event horizon. 
Thus, every time a black hole captures which-path information, the change in its entropy faces an inevitable fluctuation.

As a corollary, the trade-off relation \eqref{eq:BH TUR} leads to several important consequences. 
First, it can be recast as an uncertainty relation by using \eqref{eq:number variance ineq}, 
\begin{align}
    \Var[\hat N]
    \dfrac{
        \mathbb V
        \kagikako{
            \Delta \delta S
        }
    }
    {
        \ave{ \Delta \delta S }^2
    }
    \geq 
        \dfrac{1}{4}\,,
\end{align}
which suggests that the precision of the entropy response is limited by photon number fluctuations.

The black hole entropy trade-off relation \eqref{eq:BH TUR} can also be turned into a bound on the fluctuations of the horizon area. 
Restricting to the Bekenstein-Hawking contribution to the entropy, $\delta S \approx \delta A[\mathcal C]/(4G\ts{N})$, with $A[\mathcal C]$ the area of a horizon cross section, we find $\sqrt{\mathbb V[ \Delta \delta A ]} \geq 2\pi G\ts{N} \braket{\renorm{\hat Q[\xi]}}/(\kappa \sqrt{\braket{\hat N}})$. 
For a $(3+1)$-dimensional Schwarzschild black hole, one has the horizon radius $r\ts{h}=2G\ts{N}M$ and the surface gravity $\kappa=(2r\ts{h})^{-1}$. 
If Alice is hovering outside a black hole at a fixed proper distance $D$ from the horizon and conducts the DSW experiment with a charge $q$ in a spatial superposition with distance $l$ of two well-separated wave packets for some finite time $T$, the entangling photon number is known to be $\braket{\hat N}\sim r\ts{h}^3 q^2 l^2 T/D^6$ \cite{DSW.BH.2022}. 
Taking $D\sim T\sim r\ts{h}$ together with $\braket{\renorm{\hat Q[\xi]}} \gtrsim 1/r\ts{h}$ for the energy of the infalling photons, the trade-off relation reads 
\begin{align}
    \sqrt{\mathbb V[ \Delta \delta A ]}
    \gtrsim
        (\ell\ts{P}/ql)\times r\ts{h}\ell\ts{P}\,,
\end{align}
where $\ell\ts{P} = \sqrt{G\ts{N}}$ is the Planck length in units $c=\hbar =1$ and $ql$ is the effective dipole moment of the spatially superposed charge. 
The scale $r\ts{h}\ell\ts{P}$ commonly appears in studies of fluctuations in gravitational systems \cite{Parikh.area.uncertainty.2025, Ciambelli.area.fluc.2025, Sorkin.wrinkled.1996, Marolf.quantum.width.2005, Bousso.Islands.2024, Banks.Breakdown.2024} (see also \cite{Casher.BHfluc.1997, BLHu.fluctuations.2007, Thompson.Enhanced.2008}).

Finally, the lower bound in \eqref{eq:BH TUR} can be expressed in terms of the quantum mutual information $I\in [0,2]$ between the superposed charged particle and radiated photons, which quantifies the total amount of correlations between them. 
Setting $d=2$ and $\ket{\psi\ts{sys,0}}=(\ket{\mathrm L} + \ket{\mathrm R})/\sqrt{2}$, where $\ket{\mathrm L}$ and $\ket{\mathrm R}$ correspond to left and right branch of the spatially superposed charge in the DSW mechanism, the quantum mutual information after the interaction reads $I=-2 \sum_{j=\pm} p_j \log_2 p_j$, where $p_\pm \equiv (1\pm e^{-\braket{\hat N}/2})/2$. 
By using bounds $I \geq 2 (1- e^{-\braket{\hat N}})$ and $e^{-\braket{\hat N}} < 1/\braket{\hat N}$ for $\braket{\hat N} > 0$, we find a nontrivial bound: 
\begin{align}
    \dfrac{\mathbb V[\Delta \delta S]}{ \mathbb E[\Delta \delta S]^2 }
    > 
    \dfrac{1}{4} 
    \kako{
        1 - \dfrac{I}{2}
    }
    \quad (\geq 0)\,.
\end{align}
The lower bound monotonically decreases to 0 as the black hole captures more which-path information, $I\to 2$ (equivalently, $\braket{\hat N}\to \infty$). 
This means that the black hole's complete knowledge of which-path information potentially minimizes the entropy fluctuation, though it takes an infinite amount of time.

\emph{Conclusion}---%
Black holes inevitably destroy quantum coherence outside the horizon, thereby capturing photons carrying which-path information. 
We found an explicit form of the lower bound \eqref{eq:main result for charge} on the fluctuations in the infalling energy and analyzed the resulting perturbation of the black hole geometry in the stochastic semiclassical gravity framework. 
Consequently, we derived a trade-off relation for the entropy change of the black hole \eqref{eq:BH TUR} induced by the fluctuating infalling entangling photons. 
We regard this trade-off relation for black hole entropy as a step toward understanding stochastic aspects of black hole thermodynamics.


\emph{Acknowledgments}---%
K.G.-Y. thanks Tomoya Hirotani, Kuan-Nan Lin, Akira Matsumura, Yasusada Nambu, Yuki Osawa, and Erickson Tjoa for helpful discussions. 
K.G.-Y. is supported by Grant-in-Aid for JSPS Fellows Grant No. JP25KJ0048. 
Y.H. is supported by JSPS KAKENHI Grant Numbers JP23K24915 and JP24K03008.

\emph{Data availability}---%
No data were created or analyzed in this study.

\bibliography{ref}

\begin{thebibliography}{42}%
\makeatletter
\providecommand \@ifxundefined [1]{%
 \@ifx{#1\undefined}
}%
\providecommand \@ifnum [1]{%
 \ifnum #1\expandafter \@firstoftwo
 \else \expandafter \@secondoftwo
 \fi
}%
\providecommand \@ifx [1]{%
 \ifx #1\expandafter \@firstoftwo
 \else \expandafter \@secondoftwo
 \fi
}%
\providecommand \natexlab [1]{#1}%
\providecommand \enquote  [1]{``#1''}%
\providecommand \bibnamefont  [1]{#1}%
\providecommand \bibfnamefont [1]{#1}%
\providecommand \citenamefont [1]{#1}%
\providecommand \href@noop [0]{\@secondoftwo}%
\providecommand \href [0]{\begingroup \@sanitize@url \@href}%
\providecommand \@href[1]{\@@startlink{#1}\@@href}%
\providecommand \@@href[1]{\endgroup#1\@@endlink}%
\providecommand \@sanitize@url [0]{\catcode `\\12\catcode `\$12\catcode `\&12\catcode `\#12\catcode `\^12\catcode `\_12\catcode `\%12\relax}%
\providecommand \@@startlink[1]{}%
\providecommand \@@endlink[0]{}%
\providecommand \url  [0]{\begingroup\@sanitize@url \@url }%
\providecommand \@url [1]{\endgroup\@href {#1}{\urlprefix }}%
\providecommand \urlprefix  [0]{URL }%
\providecommand \Eprint [0]{\href }%
\providecommand \doibase [0]{https://doi.org/}%
\providecommand \selectlanguage [0]{\@gobble}%
\providecommand \bibinfo  [0]{\@secondoftwo}%
\providecommand \bibfield  [0]{\@secondoftwo}%
\providecommand \translation [1]{[#1]}%
\providecommand \BibitemOpen [0]{}%
\providecommand \bibitemStop [0]{}%
\providecommand \bibitemNoStop [0]{.\EOS\space}%
\providecommand \EOS [0]{\spacefactor3000\relax}%
\providecommand \BibitemShut  [1]{\csname bibitem#1\endcsname}%
\let\auto@bib@innerbib\@empty
\bibitem [{\citenamefont {Bekenstein}(1972)}]{Bekenstein.BH2ndLaw.1972}%
  \BibitemOpen
  \bibfield  {author} {\bibinfo {author} {\bibfnamefont {J.~D.}\ \bibnamefont {Bekenstein}},\ }\bibfield  {title} {\bibinfo {title} {{Black holes and the second law}},\ }\href {https://doi.org/10.1007/BF02757029} {\bibfield  {journal} {\bibinfo  {journal} {Lett. Nuovo Cimento}\ }\textbf {\bibinfo {volume} {4}},\ \bibinfo {pages} {737} (\bibinfo {year} {1972})}\BibitemShut {NoStop}%
\bibitem [{\citenamefont {Bekenstein}(1973)}]{Bekenstein.BHentropy.1973}%
  \BibitemOpen
  \bibfield  {author} {\bibinfo {author} {\bibfnamefont {J.~D.}\ \bibnamefont {Bekenstein}},\ }\bibfield  {title} {\bibinfo {title} {{Black Holes and Entropy}},\ }\href {https://doi.org/10.1103/PhysRevD.7.2333} {\bibfield  {journal} {\bibinfo  {journal} {Phys. Rev. D}\ }\textbf {\bibinfo {volume} {7}},\ \bibinfo {pages} {2333} (\bibinfo {year} {1973})}\BibitemShut {NoStop}%
\bibitem [{\citenamefont {Bekenstein}(1974)}]{Bekenstein.GSL.1974}%
  \BibitemOpen
  \bibfield  {author} {\bibinfo {author} {\bibfnamefont {J.~D.}\ \bibnamefont {Bekenstein}},\ }\bibfield  {title} {\bibinfo {title} {{Generalized second law of thermodynamics in black-hole physics}},\ }\href {https://doi.org/10.1103/PhysRevD.9.3292} {\bibfield  {journal} {\bibinfo  {journal} {Phys. Rev. D}\ }\textbf {\bibinfo {volume} {9}},\ \bibinfo {pages} {3292} (\bibinfo {year} {1974})}\BibitemShut {NoStop}%
\bibitem [{\citenamefont {Wald}(1994)}]{wald1994QFT}%
  \BibitemOpen
  \bibfield  {author} {\bibinfo {author} {\bibfnamefont {R.~M.}\ \bibnamefont {Wald}},\ }\href@noop {} {\emph {\bibinfo {title} {Quantum field theory in curved spacetime and black hole thermodynamics}}}\ (\bibinfo  {publisher} {University of Chicago Press},\ \bibinfo {year} {1994})\BibitemShut {NoStop}%
\bibitem [{\citenamefont {Gao}\ and\ \citenamefont {Wald}(2001)}]{Sijie.physical.process.2001}%
  \BibitemOpen
  \bibfield  {author} {\bibinfo {author} {\bibfnamefont {S.}~\bibnamefont {Gao}}\ and\ \bibinfo {author} {\bibfnamefont {R.~M.}\ \bibnamefont {Wald}},\ }\bibfield  {title} {\bibinfo {title} {{``Physical process version'' of the first law and the generalized second law for charged and rotating black holes}},\ }\href {https://doi.org/10.1103/PhysRevD.64.084020} {\bibfield  {journal} {\bibinfo  {journal} {Phys. Rev. D}\ }\textbf {\bibinfo {volume} {64}},\ \bibinfo {pages} {084020} (\bibinfo {year} {2001})}\BibitemShut {NoStop}%
\bibitem [{\citenamefont {Hollands}\ \emph {et~al.}(2024)\citenamefont {Hollands}, \citenamefont {Wald},\ and\ \citenamefont {Zhang}}]{Hollands.entropyBH.2024}%
  \BibitemOpen
  \bibfield  {author} {\bibinfo {author} {\bibfnamefont {S.}~\bibnamefont {Hollands}}, \bibinfo {author} {\bibfnamefont {R.~M.}\ \bibnamefont {Wald}},\ and\ \bibinfo {author} {\bibfnamefont {V.~G.}\ \bibnamefont {Zhang}},\ }\bibfield  {title} {\bibinfo {title} {{Entropy of dynamical black holes}},\ }\href {https://doi.org/10.1103/PhysRevD.110.024070} {\bibfield  {journal} {\bibinfo  {journal} {Phys. Rev. D}\ }\textbf {\bibinfo {volume} {110}},\ \bibinfo {pages} {024070} (\bibinfo {year} {2024})}\BibitemShut {NoStop}%
\bibitem [{\citenamefont {`t~Hooft}(1993)}]{tHooft.dimensional.1993}%
  \BibitemOpen
  \bibfield  {author} {\bibinfo {author} {\bibfnamefont {G.}~\bibnamefont {`t~Hooft}},\ }\bibfield  {title} {\bibinfo {title} {{Dimensional reduction in quantum gravity}},\ }\href@noop {} {\bibfield  {journal} {\bibinfo  {journal} {Conf. Proc. C}\ }\textbf {\bibinfo {volume} {930308}},\ \bibinfo {pages} {284} (\bibinfo {year} {1993})},\ \Eprint {https://arxiv.org/abs/arXiv:gr-qc/9310026} {arXiv:gr-qc/9310026} \BibitemShut {NoStop}%
\bibitem [{\citenamefont {Susskind}(1995)}]{Susskind.hologram.1995}%
  \BibitemOpen
  \bibfield  {author} {\bibinfo {author} {\bibfnamefont {L.}~\bibnamefont {Susskind}},\ }\bibfield  {title} {\bibinfo {title} {{The world as a hologram}},\ }\href {https://doi.org/10.1063/1.531249} {\bibfield  {journal} {\bibinfo  {journal} {J. Math. Phys.}\ }\textbf {\bibinfo {volume} {36}},\ \bibinfo {pages} {6377} (\bibinfo {year} {1995})}\BibitemShut {NoStop}%
\bibitem [{\citenamefont {Susskind}\ \emph {et~al.}(1993)\citenamefont {Susskind}, \citenamefont {Thorlacius},\ and\ \citenamefont {Uglum}}]{Susskind.BHcomplementarity.1993}%
  \BibitemOpen
  \bibfield  {author} {\bibinfo {author} {\bibfnamefont {L.}~\bibnamefont {Susskind}}, \bibinfo {author} {\bibfnamefont {L.}~\bibnamefont {Thorlacius}},\ and\ \bibinfo {author} {\bibfnamefont {J.}~\bibnamefont {Uglum}},\ }\bibfield  {title} {\bibinfo {title} {{The stretched horizon and black hole complementarity}},\ }\href {https://doi.org/10.1103/PhysRevD.48.3743} {\bibfield  {journal} {\bibinfo  {journal} {Phys. Rev. D}\ }\textbf {\bibinfo {volume} {48}},\ \bibinfo {pages} {3743} (\bibinfo {year} {1993})}\BibitemShut {NoStop}%
\bibitem [{\citenamefont {Hayden}\ and\ \citenamefont {Preskill}(2007)}]{Hayden.Preskill.2007}%
  \BibitemOpen
  \bibfield  {author} {\bibinfo {author} {\bibfnamefont {P.}~\bibnamefont {Hayden}}\ and\ \bibinfo {author} {\bibfnamefont {J.}~\bibnamefont {Preskill}},\ }\bibfield  {title} {\bibinfo {title} {{Black holes as mirrors: quantum information in random subsystems}},\ }\href {https://doi.org/10.1088/1126-6708/2007/09/120} {\bibfield  {journal} {\bibinfo  {journal} {Journal of High Energy Physics}\ }\textbf {\bibinfo {volume} {09}},\ \bibinfo {pages} {120} (\bibinfo {year} {2007})}\BibitemShut {NoStop}%
\bibitem [{\citenamefont {Almheiri}\ \emph {et~al.}(2013)\citenamefont {Almheiri}, \citenamefont {Marolf}, \citenamefont {Polchinski},\ and\ \citenamefont {Sully}}]{AMPS.2013}%
  \BibitemOpen
  \bibfield  {author} {\bibinfo {author} {\bibfnamefont {A.}~\bibnamefont {Almheiri}}, \bibinfo {author} {\bibfnamefont {D.}~\bibnamefont {Marolf}}, \bibinfo {author} {\bibfnamefont {J.}~\bibnamefont {Polchinski}},\ and\ \bibinfo {author} {\bibfnamefont {J.}~\bibnamefont {Sully}},\ }\bibfield  {title} {\bibinfo {title} {{Black holes: complementarity or firewalls?}},\ }\href {https://doi.org/10.1007/JHEP02(2013)062} {\bibfield  {journal} {\bibinfo  {journal} {Journal of High Energy Physics}\ }\textbf {\bibinfo {volume} {02}},\ \bibinfo {pages} {062} (\bibinfo {year} {2013})}\BibitemShut {NoStop}%
\bibitem [{\citenamefont {Hu}\ and\ \citenamefont {Verdaguer}(2003)}]{Hu.stochastic.primer.2003}%
  \BibitemOpen
  \bibfield  {author} {\bibinfo {author} {\bibfnamefont {B.~L.}\ \bibnamefont {Hu}}\ and\ \bibinfo {author} {\bibfnamefont {E.}~\bibnamefont {Verdaguer}},\ }\bibfield  {title} {\bibinfo {title} {Stochastic gravity: a primer with applications},\ }\href {https://doi.org/10.1088/0264-9381/20/6/201} {\bibfield  {journal} {\bibinfo  {journal} {Classical and Quantum Gravity}\ }\textbf {\bibinfo {volume} {20}},\ \bibinfo {pages} {R1} (\bibinfo {year} {2003})}\BibitemShut {NoStop}%
\bibitem [{\citenamefont {Hu}\ and\ \citenamefont {Verdaguer}(2008)}]{Hu.StochasticGravity.2008}%
  \BibitemOpen
  \bibfield  {author} {\bibinfo {author} {\bibfnamefont {B.~L.}\ \bibnamefont {Hu}}\ and\ \bibinfo {author} {\bibfnamefont {E.}~\bibnamefont {Verdaguer}},\ }\bibfield  {title} {\bibinfo {title} {{Stochastic Gravity: Theory and Applications}},\ }\href {https://doi.org/10.12942/lrr-2008-3} {\bibfield  {journal} {\bibinfo  {journal} {Living Rev. Relativ.}\ }\textbf {\bibinfo {volume} {11}},\ \bibinfo {pages} {3} (\bibinfo {year} {2008})}\BibitemShut {NoStop}%
\bibitem [{\citenamefont {Danielson}\ \emph {et~al.}(2022)\citenamefont {Danielson}, \citenamefont {Satishchandran},\ and\ \citenamefont {Wald}}]{DSW.BH.2022}%
  \BibitemOpen
  \bibfield  {author} {\bibinfo {author} {\bibfnamefont {D.~L.}\ \bibnamefont {Danielson}}, \bibinfo {author} {\bibfnamefont {G.}~\bibnamefont {Satishchandran}},\ and\ \bibinfo {author} {\bibfnamefont {R.~M.}\ \bibnamefont {Wald}},\ }\bibfield  {title} {\bibinfo {title} {{Black holes decohere quantum superpositions}},\ }\href {https://doi.org/10.1142/S0218271822410036} {\bibfield  {journal} {\bibinfo  {journal} {International Journal of Modern Physics D}\ }\textbf {\bibinfo {volume} {31}},\ \bibinfo {pages} {2241003} (\bibinfo {year} {2022})}\BibitemShut {NoStop}%
\bibitem [{\citenamefont {Danielson}\ \emph {et~al.}(2023)\citenamefont {Danielson}, \citenamefont {Satishchandran},\ and\ \citenamefont {Wald}}]{DSW.Killing.2023}%
  \BibitemOpen
  \bibfield  {author} {\bibinfo {author} {\bibfnamefont {D.~L.}\ \bibnamefont {Danielson}}, \bibinfo {author} {\bibfnamefont {G.}~\bibnamefont {Satishchandran}},\ and\ \bibinfo {author} {\bibfnamefont {R.~M.}\ \bibnamefont {Wald}},\ }\bibfield  {title} {\bibinfo {title} {{Killing horizons decohere quantum superpositions}},\ }\href {https://doi.org/10.1103/PhysRevD.108.025007} {\bibfield  {journal} {\bibinfo  {journal} {Phys. Rev. D}\ }\textbf {\bibinfo {volume} {108}},\ \bibinfo {pages} {025007} (\bibinfo {year} {2023})}\BibitemShut {NoStop}%
\bibitem [{\citenamefont {Gralla}\ and\ \citenamefont {Wei}(2024)}]{Gralla.rotatingBH.2024}%
  \BibitemOpen
  \bibfield  {author} {\bibinfo {author} {\bibfnamefont {S.~E.}\ \bibnamefont {Gralla}}\ and\ \bibinfo {author} {\bibfnamefont {H.}~\bibnamefont {Wei}},\ }\bibfield  {title} {\bibinfo {title} {{Decoherence from horizons: General formulation and rotating black holes}},\ }\href {https://doi.org/10.1103/PhysRevD.109.065031} {\bibfield  {journal} {\bibinfo  {journal} {Phys. Rev. D}\ }\textbf {\bibinfo {volume} {109}},\ \bibinfo {pages} {065031} (\bibinfo {year} {2024})}\BibitemShut {NoStop}%
\bibitem [{\citenamefont {Danielson}\ \emph {et~al.}(2025{\natexlab{a}})\citenamefont {Danielson}, \citenamefont {Satishchandran},\ and\ \citenamefont {Wald}}]{DSW.Local.2025}%
  \BibitemOpen
  \bibfield  {author} {\bibinfo {author} {\bibfnamefont {D.~L.}\ \bibnamefont {Danielson}}, \bibinfo {author} {\bibfnamefont {G.}~\bibnamefont {Satishchandran}},\ and\ \bibinfo {author} {\bibfnamefont {R.~M.}\ \bibnamefont {Wald}},\ }\bibfield  {title} {\bibinfo {title} {{Local description of decoherence of quantum superpositions by black holes and other bodies}},\ }\href {https://doi.org/10.1103/PhysRevD.111.025014} {\bibfield  {journal} {\bibinfo  {journal} {Phys. Rev. D}\ }\textbf {\bibinfo {volume} {111}},\ \bibinfo {pages} {025014} (\bibinfo {year} {2025}{\natexlab{a}})}\BibitemShut {NoStop}%
\bibitem [{\citenamefont {Li}(2025)}]{Li.RNBH.2025}%
  \BibitemOpen
  \bibfield  {author} {\bibinfo {author} {\bibfnamefont {R.}~\bibnamefont {Li}},\ }\bibfield  {title} {\bibinfo {title} {{Decoherence of quantum superpositions by Reissner-Nordstr\"om black holes}},\ }\href {https://doi.org/10.1103/PhysRevD.111.024040} {\bibfield  {journal} {\bibinfo  {journal} {Phys. Rev. D}\ }\textbf {\bibinfo {volume} {111}},\ \bibinfo {pages} {024040} (\bibinfo {year} {2025})}\BibitemShut {NoStop}%
\bibitem [{\citenamefont {Danielson}\ \emph {et~al.}(2025{\natexlab{b}})\citenamefont {Danielson}, \citenamefont {Kudler-Flam}, \citenamefont {Satishchandran},\ and\ \citenamefont {Wald}}]{DKSW.minimize.2025}%
  \BibitemOpen
  \bibfield  {author} {\bibinfo {author} {\bibfnamefont {D.~L.}\ \bibnamefont {Danielson}}, \bibinfo {author} {\bibfnamefont {J.}~\bibnamefont {Kudler-Flam}}, \bibinfo {author} {\bibfnamefont {G.}~\bibnamefont {Satishchandran}},\ and\ \bibinfo {author} {\bibfnamefont {R.~M.}\ \bibnamefont {Wald}},\ }\bibfield  {title} {\bibinfo {title} {{How to minimize the decoherence caused by black holes}},\ }\href {https://doi.org/10.1103/67vv-km43} {\bibfield  {journal} {\bibinfo  {journal} {Phys. Rev. D}\ }\textbf {\bibinfo {volume} {112}},\ \bibinfo {pages} {025012} (\bibinfo {year} {2025}{\natexlab{b}})}\BibitemShut {NoStop}%
\bibitem [{\citenamefont {Wilson-Gerow}\ \emph {et~al.}(2024)\citenamefont {Wilson-Gerow}, \citenamefont {Dugad},\ and\ \citenamefont {Chen}}]{WilsonGerow.warmHorizon.2024}%
  \BibitemOpen
  \bibfield  {author} {\bibinfo {author} {\bibfnamefont {J.}~\bibnamefont {Wilson-Gerow}}, \bibinfo {author} {\bibfnamefont {A.}~\bibnamefont {Dugad}},\ and\ \bibinfo {author} {\bibfnamefont {Y.}~\bibnamefont {Chen}},\ }\bibfield  {title} {\bibinfo {title} {{Decoherence by warm horizons}},\ }\href {https://doi.org/10.1103/PhysRevD.110.045002} {\bibfield  {journal} {\bibinfo  {journal} {Phys. Rev. D}\ }\textbf {\bibinfo {volume} {110}},\ \bibinfo {pages} {045002} (\bibinfo {year} {2024})}\BibitemShut {NoStop}%
\bibitem [{\citenamefont {Tjoa}(2023)}]{Tjoa.Nonperturbative.2023}%
  \BibitemOpen
  \bibfield  {author} {\bibinfo {author} {\bibfnamefont {E.}~\bibnamefont {Tjoa}},\ }\bibfield  {title} {\bibinfo {title} {{Nonperturbative simple-generated interactions with a quantum field for arbitrary Gaussian states}},\ }\href {https://doi.org/10.1103/PhysRevD.108.045003} {\bibfield  {journal} {\bibinfo  {journal} {Phys. Rev. D}\ }\textbf {\bibinfo {volume} {108}},\ \bibinfo {pages} {045003} (\bibinfo {year} {2023})}\BibitemShut {NoStop}%
\bibitem [{\citenamefont {Blanes}\ \emph {et~al.}(2009)\citenamefont {Blanes}, \citenamefont {Casas}, \citenamefont {Oteo},\ and\ \citenamefont {Ros}}]{Blanes.Magnus.2009}%
  \BibitemOpen
  \bibfield  {author} {\bibinfo {author} {\bibfnamefont {S.}~\bibnamefont {Blanes}}, \bibinfo {author} {\bibfnamefont {F.}~\bibnamefont {Casas}}, \bibinfo {author} {\bibfnamefont {J.}~\bibnamefont {Oteo}},\ and\ \bibinfo {author} {\bibfnamefont {J.}~\bibnamefont {Ros}},\ }\bibfield  {title} {\bibinfo {title} {{The Magnus expansion and some of its applications}},\ }\href {https://doi.org/https://doi.org/10.1016/j.physrep.2008.11.001} {\bibfield  {journal} {\bibinfo  {journal} {Physics Reports}\ }\textbf {\bibinfo {volume} {470}},\ \bibinfo {pages} {151} (\bibinfo {year} {2009})}\BibitemShut {NoStop}%
\bibitem [{\citenamefont {Landulfo}(2016)}]{Landulfo2016magnus1}%
  \BibitemOpen
  \bibfield  {author} {\bibinfo {author} {\bibfnamefont {A.~G.~S.}\ \bibnamefont {Landulfo}},\ }\bibfield  {title} {\bibinfo {title} {Nonperturbative approach to relativistic quantum communication channels},\ }\href {https://doi.org/10.1103/PhysRevD.93.104019} {\bibfield  {journal} {\bibinfo  {journal} {Phys. Rev. D}\ }\textbf {\bibinfo {volume} {93}},\ \bibinfo {pages} {104019} (\bibinfo {year} {2016})}\BibitemShut {NoStop}%
\bibitem [{\citenamefont {Cozzella}\ \emph {et~al.}(2020)\citenamefont {Cozzella}, \citenamefont {Fulling}, \citenamefont {Landulfo},\ and\ \citenamefont {Matsas}}]{Cozzella.UDW.2020}%
  \BibitemOpen
  \bibfield  {author} {\bibinfo {author} {\bibfnamefont {G.}~\bibnamefont {Cozzella}}, \bibinfo {author} {\bibfnamefont {S.~A.}\ \bibnamefont {Fulling}}, \bibinfo {author} {\bibfnamefont {A.~G.~S.}\ \bibnamefont {Landulfo}},\ and\ \bibinfo {author} {\bibfnamefont {G.~E.~A.}\ \bibnamefont {Matsas}},\ }\bibfield  {title} {\bibinfo {title} {{Uniformly accelerated classical sources as limits of Unruh-DeWitt detectors}},\ }\href {https://doi.org/10.1103/PhysRevD.102.105016} {\bibfield  {journal} {\bibinfo  {journal} {Phys. Rev. D}\ }\textbf {\bibinfo {volume} {102}},\ \bibinfo {pages} {105016} (\bibinfo {year} {2020})}\BibitemShut {NoStop}%
\bibitem [{\citenamefont {Gallock-Yoshimura}\ \emph {et~al.}(2025)\citenamefont {Gallock-Yoshimura}, \citenamefont {Osawa},\ and\ \citenamefont {Nambu}}]{Gallock.Acceleration.2025}%
  \BibitemOpen
  \bibfield  {author} {\bibinfo {author} {\bibfnamefont {K.}~\bibnamefont {Gallock-Yoshimura}}, \bibinfo {author} {\bibfnamefont {Y.}~\bibnamefont {Osawa}},\ and\ \bibinfo {author} {\bibfnamefont {Y.}~\bibnamefont {Nambu}},\ }\bibfield  {title} {\bibinfo {title} {{Acceleration-induced radiation from a qudit particle detector model}},\ }\href {https://doi.org/10.1103/PhysRevD.111.105012} {\bibfield  {journal} {\bibinfo  {journal} {Phys. Rev. D}\ }\textbf {\bibinfo {volume} {111}},\ \bibinfo {pages} {105012} (\bibinfo {year} {2025})}\BibitemShut {NoStop}%
\bibitem [{\citenamefont {Barato}\ and\ \citenamefont {Seifert}(2015)}]{Barato.TUR.2015}%
  \BibitemOpen
  \bibfield  {author} {\bibinfo {author} {\bibfnamefont {A.~C.}\ \bibnamefont {Barato}}\ and\ \bibinfo {author} {\bibfnamefont {U.}~\bibnamefont {Seifert}},\ }\bibfield  {title} {\bibinfo {title} {{Thermodynamic Uncertainty Relation for Biomolecular Processes}},\ }\href {https://doi.org/10.1103/PhysRevLett.114.158101} {\bibfield  {journal} {\bibinfo  {journal} {Phys. Rev. Lett.}\ }\textbf {\bibinfo {volume} {114}},\ \bibinfo {pages} {158101} (\bibinfo {year} {2015})}\BibitemShut {NoStop}%
\bibitem [{\citenamefont {Gingrich}\ \emph {et~al.}(2016)\citenamefont {Gingrich}, \citenamefont {Horowitz}, \citenamefont {Perunov},\ and\ \citenamefont {England}}]{Gingrich.Dissipation.2016}%
  \BibitemOpen
  \bibfield  {author} {\bibinfo {author} {\bibfnamefont {T.~R.}\ \bibnamefont {Gingrich}}, \bibinfo {author} {\bibfnamefont {J.~M.}\ \bibnamefont {Horowitz}}, \bibinfo {author} {\bibfnamefont {N.}~\bibnamefont {Perunov}},\ and\ \bibinfo {author} {\bibfnamefont {J.~L.}\ \bibnamefont {England}},\ }\bibfield  {title} {\bibinfo {title} {{Dissipation Bounds All Steady-State Current Fluctuations}},\ }\href {https://doi.org/10.1103/PhysRevLett.116.120601} {\bibfield  {journal} {\bibinfo  {journal} {Phys. Rev. Lett.}\ }\textbf {\bibinfo {volume} {116}},\ \bibinfo {pages} {120601} (\bibinfo {year} {2016})}\BibitemShut {NoStop}%
\bibitem [{\citenamefont {Hasegawa}(2020)}]{Hasegawa.QTUR.measurement.2020}%
  \BibitemOpen
  \bibfield  {author} {\bibinfo {author} {\bibfnamefont {Y.}~\bibnamefont {Hasegawa}},\ }\bibfield  {title} {\bibinfo {title} {{Quantum Thermodynamic Uncertainty Relation for Continuous Measurement}},\ }\href {https://doi.org/10.1103/PhysRevLett.125.050601} {\bibfield  {journal} {\bibinfo  {journal} {Phys. Rev. Lett.}\ }\textbf {\bibinfo {volume} {125}},\ \bibinfo {pages} {050601} (\bibinfo {year} {2020})}\BibitemShut {NoStop}%
\bibitem [{\citenamefont {Hasegawa}(2021)}]{Hasegawa.QTURgeneral.2021}%
  \BibitemOpen
  \bibfield  {author} {\bibinfo {author} {\bibfnamefont {Y.}~\bibnamefont {Hasegawa}},\ }\bibfield  {title} {\bibinfo {title} {{Thermodynamic Uncertainty Relation for General Open Quantum Systems}},\ }\href {https://doi.org/10.1103/PhysRevLett.126.010602} {\bibfield  {journal} {\bibinfo  {journal} {Phys. Rev. Lett.}\ }\textbf {\bibinfo {volume} {126}},\ \bibinfo {pages} {010602} (\bibinfo {year} {2021})}\BibitemShut {NoStop}%
\bibitem [{\citenamefont {Iyer}\ and\ \citenamefont {Wald}(1994)}]{Iyer.Noether.1994}%
  \BibitemOpen
  \bibfield  {author} {\bibinfo {author} {\bibfnamefont {V.}~\bibnamefont {Iyer}}\ and\ \bibinfo {author} {\bibfnamefont {R.~M.}\ \bibnamefont {Wald}},\ }\bibfield  {title} {\bibinfo {title} {{Some properties of the Noether charge and a proposal for dynamical black hole entropy}},\ }\href {https://doi.org/10.1103/PhysRevD.50.846} {\bibfield  {journal} {\bibinfo  {journal} {Phys. Rev. D}\ }\textbf {\bibinfo {volume} {50}},\ \bibinfo {pages} {846} (\bibinfo {year} {1994})}\BibitemShut {NoStop}%
\bibitem [{\citenamefont {Wald}\ and\ \citenamefont {Zoupas}(2000)}]{Wald.Zoupas.2000}%
  \BibitemOpen
  \bibfield  {author} {\bibinfo {author} {\bibfnamefont {R.~M.}\ \bibnamefont {Wald}}\ and\ \bibinfo {author} {\bibfnamefont {A.}~\bibnamefont {Zoupas}},\ }\bibfield  {title} {\bibinfo {title} {{General definition of ``conserved quantities'' in general relativity and other theories of gravity}},\ }\href {https://doi.org/10.1103/PhysRevD.61.084027} {\bibfield  {journal} {\bibinfo  {journal} {Phys. Rev. D}\ }\textbf {\bibinfo {volume} {61}},\ \bibinfo {pages} {084027} (\bibinfo {year} {2000})}\BibitemShut {NoStop}%
\bibitem [{\citenamefont {Hollands}\ and\ \citenamefont {Wald}(2012)}]{Hollands.stability.2012}%
  \BibitemOpen
  \bibfield  {author} {\bibinfo {author} {\bibfnamefont {S.}~\bibnamefont {Hollands}}\ and\ \bibinfo {author} {\bibfnamefont {R.~M.}\ \bibnamefont {Wald}},\ }\bibfield  {title} {\bibinfo {title} {{Stability of Black Holes and Black Branes}},\ }\href {https://doi.org/10.1007/s00220-012-1638-1} {\bibfield  {journal} {\bibinfo  {journal} {Communications in Mathematical Physics}\ }\textbf {\bibinfo {volume} {321}},\ \bibinfo {pages} {629^^e2^^80^^93680} (\bibinfo {year} {2012})}\BibitemShut {NoStop}%
\bibitem [{Sup()}]{Supplemental}%
  \BibitemOpen
  \href@noop {} {}\bibinfo {note} {See Supplemental Material for details.}\BibitemShut {Stop}%
\bibitem [{\citenamefont {Parikh}\ and\ \citenamefont {Pereira}(2025)}]{Parikh.area.uncertainty.2025}%
  \BibitemOpen
  \bibfield  {author} {\bibinfo {author} {\bibfnamefont {M.}~\bibnamefont {Parikh}}\ and\ \bibinfo {author} {\bibfnamefont {J.}~\bibnamefont {Pereira}},\ }\bibfield  {title} {\bibinfo {title} {{Quantum uncertainty in the area of a black hole}},\ }\href {https://doi.org/10.1007/JHEP09(2025)137} {\bibfield  {journal} {\bibinfo  {journal} {J. High Energy Phys.}\ }\textbf {\bibinfo {volume} {09}}\bibinfo  {number} { (2025)},\ \bibinfo {pages} {137}}\BibitemShut {NoStop}%
\bibitem [{\citenamefont {Ciambelli}\ \emph {et~al.}(2025)\citenamefont {Ciambelli}, \citenamefont {He},\ and\ \citenamefont {Zurek}}]{Ciambelli.area.fluc.2025}%
  \BibitemOpen
\bibfield  {number} {  }\bibfield  {author} {\bibinfo {author} {\bibfnamefont {L.}~\bibnamefont {Ciambelli}}, \bibinfo {author} {\bibfnamefont {T.}~\bibnamefont {He}},\ and\ \bibinfo {author} {\bibfnamefont {K.~M.}\ \bibnamefont {Zurek}},\ }\bibfield  {title} {\bibinfo {title} {{Quantum area fluctuations from gravitational phase space}},\ }\href {https://doi.org/10.1007/JHEP08(2025)199} {\bibfield  {journal} {\bibinfo  {journal} {J. High Energy Phys.}\ }\textbf {\bibinfo {volume} {08}}\bibinfo  {number} { (2025)},\ \bibinfo {pages} {199}}\BibitemShut {NoStop}%
\bibitem [{\citenamefont {Sorkin}(1996)}]{Sorkin.wrinkled.1996}%
  \BibitemOpen
\bibfield  {number} {  }\bibfield  {author} {\bibinfo {author} {\bibfnamefont {R.~D.}\ \bibnamefont {Sorkin}},\ }\bibfield  {title} {\bibinfo {title} {{How Wrinkled is the Surface of a Black Hole?}},\ }in\ \href@noop {} {\emph {\bibinfo {booktitle} {Proceedings of the First Australasian Conference on General Relativity and Gravitation}}},\ \bibinfo {editor} {edited by\ \bibinfo {editor} {\bibfnamefont {D.}~\bibnamefont {Wiltshire}}}\ (\bibinfo  {publisher} {University of Adelaide},\ \bibinfo {year} {1996})\ pp.\ \bibinfo {pages} {163--174},\ \Eprint {https://arxiv.org/abs/arXiv:gr-qc/9701056} {arXiv:gr-qc/9701056} \BibitemShut {NoStop}%
\bibitem [{\citenamefont {Marolf}(2005)}]{Marolf.quantum.width.2005}%
  \BibitemOpen
  \bibfield  {author} {\bibinfo {author} {\bibfnamefont {D.}~\bibnamefont {Marolf}},\ }\bibfield  {title} {\bibinfo {title} {{On the Quantum Width of a Black Hole Horizon}},\ }\href {https://doi.org/10.1007/3-540-26798-0_9} {\bibfield  {journal} {\bibinfo  {journal} {Springer Proc. Phys.}\ }\textbf {\bibinfo {volume} {98}},\ \bibinfo {pages} {99} (\bibinfo {year} {2005})}\BibitemShut {NoStop}%
\bibitem [{\citenamefont {Bousso}\ and\ \citenamefont {Penington}(2024)}]{Bousso.Islands.2024}%
  \BibitemOpen
  \bibfield  {author} {\bibinfo {author} {\bibfnamefont {R.}~\bibnamefont {Bousso}}\ and\ \bibinfo {author} {\bibfnamefont {G.}~\bibnamefont {Penington}},\ }\bibfield  {title} {\bibinfo {title} {{Islands far outside the horizon}},\ }\href {https://doi.org/10.1007/JHEP11(2024)164} {\bibfield  {journal} {\bibinfo  {journal} {Journal of High Energy Physics}\ }\textbf {\bibinfo {volume} {11}},\ \bibinfo {pages} {164} (\bibinfo {year} {2024})}\BibitemShut {NoStop}%
\bibitem [{\citenamefont {Banks}\ \emph {et~al.}(2024)\citenamefont {Banks}, \citenamefont {Draper},\ and\ \citenamefont {Karydas}}]{Banks.Breakdown.2024}%
  \BibitemOpen
  \bibfield  {author} {\bibinfo {author} {\bibfnamefont {T.}~\bibnamefont {Banks}}, \bibinfo {author} {\bibfnamefont {P.}~\bibnamefont {Draper}},\ and\ \bibinfo {author} {\bibfnamefont {M.}~\bibnamefont {Karydas}},\ }\bibfield  {title} {\bibinfo {title} {{Breakdown of field theory in near-horizon regions}},\ }\href {https://doi.org/10.1007/JHEP06(2024)153} {\bibfield  {journal} {\bibinfo  {journal} {Journal of High Energy Physics}\ }\textbf {\bibinfo {volume} {06}},\ \bibinfo {pages} {153} (\bibinfo {year} {2024})}\BibitemShut {NoStop}%
\bibitem [{\citenamefont {Casher}\ \emph {et~al.}(1997)\citenamefont {Casher}, \citenamefont {Englert}, \citenamefont {Itzhaki}, \citenamefont {Massar},\ and\ \citenamefont {Parentani}}]{Casher.BHfluc.1997}%
  \BibitemOpen
  \bibfield  {author} {\bibinfo {author} {\bibfnamefont {A.}~\bibnamefont {Casher}}, \bibinfo {author} {\bibfnamefont {F.}~\bibnamefont {Englert}}, \bibinfo {author} {\bibfnamefont {N.}~\bibnamefont {Itzhaki}}, \bibinfo {author} {\bibfnamefont {S.}~\bibnamefont {Massar}},\ and\ \bibinfo {author} {\bibfnamefont {R.}~\bibnamefont {Parentani}},\ }\bibfield  {title} {\bibinfo {title} {{Black hole horizon fluctuations}},\ }\href {https://doi.org/https://doi.org/10.1016/S0550-3213(96)00613-X} {\bibfield  {journal} {\bibinfo  {journal} {Nuclear Physics B}\ }\textbf {\bibinfo {volume} {484}},\ \bibinfo {pages} {419} (\bibinfo {year} {1997})}\BibitemShut {NoStop}%
\bibitem [{\citenamefont {Hu}\ and\ \citenamefont {Roura}(2007)}]{BLHu.fluctuations.2007}%
  \BibitemOpen
  \bibfield  {author} {\bibinfo {author} {\bibfnamefont {B.~L.}\ \bibnamefont {Hu}}\ and\ \bibinfo {author} {\bibfnamefont {A.}~\bibnamefont {Roura}},\ }\bibfield  {title} {\bibinfo {title} {{Metric fluctuations of an evaporating black hole from backreaction of stress tensor fluctuations}},\ }\href {https://doi.org/10.1103/PhysRevD.76.124018} {\bibfield  {journal} {\bibinfo  {journal} {Phys. Rev. D}\ }\textbf {\bibinfo {volume} {76}},\ \bibinfo {pages} {124018} (\bibinfo {year} {2007})}\BibitemShut {NoStop}%
\bibitem [{\citenamefont {Thompson}\ and\ \citenamefont {Ford}(2008)}]{Thompson.Enhanced.2008}%
  \BibitemOpen
  \bibfield  {author} {\bibinfo {author} {\bibfnamefont {R.~T.}\ \bibnamefont {Thompson}}\ and\ \bibinfo {author} {\bibfnamefont {L.~H.}\ \bibnamefont {Ford}},\ }\bibfield  {title} {\bibinfo {title} {{Enhanced black hole horizon fluctuations}},\ }\href {https://doi.org/10.1103/PhysRevD.78.024014} {\bibfield  {journal} {\bibinfo  {journal} {Phys. Rev. D}\ }\textbf {\bibinfo {volume} {78}},\ \bibinfo {pages} {024014} (\bibinfo {year} {2008})}\BibitemShut {NoStop}%
\end{thebibliography}%

\appendix
\EndMatter

\emph{Appendix: Derivation of  Eq.~\eqref{eq:main result for charge}}---%
We provide relevant calculations needed for deriving Eq.~\eqref{eq:main result for charge}. 
From the commutation relation $[\hat \phi(f), \phi(f')]=\ii E(f,f')\id$, we obtain the relation $[\hat \phi(f), \nabla_\alpha \hat \phi(\sx)]=-\ii \partial_\alpha (Ef) \id$, where $\nabla_\alpha$ is the covariant derivative at $\sx \in \mfd$ and $Ef(\sx)\coloneqq \int_\mfd \dvol{g'}\,E(\sx, \sx') f(\sx')$ is the real advanced-minus-retarded solution to the Klein-Gordon equation. 
From this commutation relation, we find $\hat U^\dag \nabla_\alpha \hat\phi \hat U=\nabla_\alpha \hat \phi + \partial_\alpha (Ef) \hat S$ for the unitary operator given in \eqref{eq:unitary operator}, and thus obtain \eqref{eq:stress tensor evolution}. 
Equation~\eqref{eq:stress tensor evolution} is used for evaluating the statistical moments of $\renorm{\hat Q[\xi]}$. 
For the quasifree Hadamard initial state, we have vanishing odd-point correlators, thus $\braket{\renorm{\hat T_{\mu \nu}}}=\braket{\hat T_{\mu \nu}} - \braket{\hat T_{\mu \nu}}_0=\braket{\hat S^2} \mathcal T_{\mu \nu}$. 
We also have 
\begin{align}
    \braket{\hat T_{\mu \nu} \hat T_{\rho \sigma}}
    &=
        \braket{\hat T_{\mu \nu} \hat T_{\rho \sigma}}_0
        +
        \braket{\hat S^2} 
        (
            \braket{\hat T_{\mu \nu}}_0 \mathcal T_{\rho \sigma}
            +
            \braket{\hat T_{\rho \sigma}}_0 \mathcal T_{\mu \nu}
        ) \notag \\
    &
        +
        \braket{\hat S^2}
        \braket{ T_{\mu \nu}[\hat \phi, Ef] T_{\rho \sigma}[\hat \phi, Ef] }_0
        +\braket{\hat S^4} \mathcal{T}_{\mu \nu} \mathcal{T}_{\rho \sigma}\,, \notag 
\end{align}
and therefore, 
\begin{align}
    &\braket{\hat T_{\mu \nu} \hat T_{\rho \sigma}}
    -
    \braket{\hat T_{\mu \nu}} \braket{\hat T_{\rho \sigma}} 
    =
        \braket{\hat T_{\mu \nu} \hat T_{\rho \sigma}}_0
        -
        \braket{\hat T_{\mu \nu}}_0 \braket{\hat T_{\rho \sigma}}_0 \notag \\
        &
        +
        \braket{\hat S^2}
        \braket{ T_{\mu \nu}[\hat \phi, Ef] T_{\rho \sigma}[\hat \phi, Ef] }_0
        +
        \Var[\hat S^2] 
        \mathcal{T}_{\mu \nu} \mathcal{T}_{\rho \sigma}\,.
\end{align}
Noticing that $\Var[\renorm{\hat Q[\xi]}]=\Var[\hat Q[\xi]]$, the direct computation of 
\begin{align}
    \Var[\hat Q[\xi]]
    &=
        \int_\Sigma \dd \Sigma^\mu 
        \int_\Sigma \dd \Sigma^\rho\,
        \xi^\nu \xi^\sigma 
        (
            \braket{\hat T_{\mu \nu} \hat T_{\rho \sigma}}
            -
            \braket{\hat T_{\mu \nu}} \braket{\hat T_{\rho \sigma}}
        ) \notag 
\end{align}
yields Eq.~\eqref{eq:varQ intermediate}.

Finally, the lower bound $\mathrm{Im} \braket{\hat \Phi(h_f) \hat \phi(f)}_0$ in the Cauchy-Schwarz inequality \eqref{eq:Cauchy-Schwarz} is computed as follows. 
By using the identities, $\ii Ef(\sx)=\braket{\hat \phi(\sx) \hat \phi(f)}_0 - \braket{\hat \phi(f) \hat \phi(\sx)}_0$ and $\overline{ \braket{\hat \phi(\sx) \hat \phi(\sx')}_0 }=\braket{\hat \phi(\sx') \hat \phi(\sx)}_0$ (where $\overline{z}$ is a complex conjugate of $z\in \C$), we have 
\begin{align}
    &2 \textrm{Im}\braket{\hat \Phi(h_f) \hat \phi(f)}_0 \notag \\
    &=
        -\ii \int_\Sigma \dd \Sigma^\mu \xi^\nu\,
        (
            \braket{ T_{\mu \nu}[\hat \phi, Ef] \hat \phi(f) }_0
            -
            \overline{
                \braket{ T_{\mu \nu}[\hat \phi, Ef] \hat \phi(f) }_0
            }
        )
        \notag \\
    &=
        \int_\Sigma \dd \Sigma^\mu\, \mathcal{T}_{\mu \nu} \xi^\nu\,.
\end{align}

\end{document}